\documentclass[12pt]{article}
\usepackage{graphicx}
\DeclareGraphicsExtensions{.pdf}
\usepackage{float} %Include figure filesusepackage{graphicx} %Include figure files
\usepackage{amsmath}
\usepackage{amsfonts}   
\usepackage{amssymb}

\begin{document}
\date{}
%%%%%%%%%%%%%%%%%%%%
\title{{\bf{\Large Holographic charge transport in non commutative gauge theories}}}
%%%%%%%%%%%%%%%%%%%%
\author{
 {\bf {\normalsize Dibakar Roychowdhury}$
$\thanks{E-mail:  dibakarphys@gmail.com, dibakar@cts.iisc.ernet.in}}\\
 {\normalsize Centre for High Energy Physics, Indian Institute of Science, }
\\{\normalsize C.V. Raman Avenue, Bangalore 560012, Karnataka, India}
%\\[0.3cm]
}
%\date{}

\maketitle
%%%%%%%%%%%%%%%%%%%%%%%%%%%%%%%%%%%%%%%%%%%%%%%%%%%%%%%%%%%%%%%%%%%%%%%%%%%%
\begin{abstract}
In this paper, based on the holographic techniques, we explore the hydrodynamics of charge diffusion phenomena in non commutative $ \mathcal{N}=4 $ SYM plasma at strong coupling. In our analysis, we compute the $ R $ charge diffusion rates both along commutative as well as the non commutative coordinates of the brane. It turns out that unlike the case for the shear viscosity, the DC conductivity along the non commutative direction of the brane differs significantly from that of its cousin corresponding to the commutative direction of the brane. Such a discrepancy however smoothly goes away in the limit of the vanishing non commutativity.
\end{abstract}

%%%%%%%%%%%%%%%%%%%%%%%%%%%%%%%%%%%%%%%%%%%%%%%%%%%%%%%%%%%%%%%%%%%
\section{Overview and Motivation}
%%%%%%%%%%%%%%%%%%%%%%%%%%%%%%%
Almost for the past one decade, the AdS/CFT correspondence \cite{Maldacena:1997re}-\cite{Aharony:1999ti} has been found to provide an extremely elegant tool in order to explore various physical properties of strongly coupled (gauge theory) plasma at sufficiently high temperatures. The hydrodynamic description of such strongly coupled gauge theories has been studied quite successfully by considering asymptotically AdS black holes in the dual gravitational counterpart \cite{Policastro:2001yc}-\cite{Kovtun:2008kx}. The underlying motivation behind such analysis rests on the fact that the Quark Gluon Plasma (QGP) produced at RHIC, Brookhaven is strongly coupled where the usual techniques of perturbative Quantum Field Theory (QFT) do not apply.

Apart from being strongly coupled, the other characteristic feature of the QGP produced at RHIC is the anisotropic expansion of the fireball during the very early stage of the collision \cite{Ryblewski:2010bs}-\cite{Martinez:2010sd} which therefore has driven a lot of attention in the context of holography \cite{Mateos:2011ix}-\cite{Cheng:2014qia}. In \cite{Mateos:2011ix}-\cite{Mateos:2011tv}, the authors had proposed a systematic anisotropic construction in the context of Einstein-axion-dilaton gravity where they have considered a particular anisotropic ($ \theta $ deformed) version of $ \mathcal{N}=4 $ SYM plasma namely, $ \delta S_{YM}\sim \int \theta(z) Tr F\wedge F $, where the $ \theta $ parameter (which is dual to the axion in the bulk) depends linearly on one of the spatial directions of the brane. The corresponding hydrodynamic analysis of their model has been performed in \cite{Rebhan:2011vd}. The key outcomes of their analysis could be summarized as follows: (1) The DC conductivity along the isotropic direction of the brane is different from that of its value corresponding to the anisotropic direction, and most importantly, (2) the shear viscosity to entropy ($ \eta/s $) ratio corresponding to the longitudinal fluctuations has been found to differ significantly from that of its value computed from the transverse fluctuations. The most significant outcome of their analysis rests on the fact that one could have a natural violation of the conjectured lower bound on $ \eta/s $ ratio solely from the anisotropic considerations even in the context of Einstein gravity \cite{Rebhan:2011vd}.

Even before these analysis had performed, in \cite{Landsteiner:2007bd} the authors had studied hydrodynamics of a strongly coupled plasma in a slightly different context of anisotropy which was driven due to presence of the non commutativity along different spatial directions of the $ Dp $ brane in the presence of a background NS B field. Holographically such theories are supposed to describe non commutative $ \mathcal{N}=4 $ SYM plasma at strong coupling \cite{Seiberg:1999vs}-\cite{Hashimoto:1999ut}. In their analysis \cite{Landsteiner:2007bd}, the authors had found that despite of the spatial anisotropy (that is caused due to the distinction between the commutative and the non commutative spatial directions) the shear viscosity to entropy ($ \eta/s $) ratio turns out to be universal for two different shear channels. The reason that the universality of the bound is still maintained in the non commutative scenario could be understood in terms of the holographic stress tensor which surprisingly turns out to be the same as that of the commutative theory \cite{Landsteiner:2007bd}.

In summary, from the comparative analysis in the previous two paragraphs, one should be able to note that the $ \theta $ deformed $ \mathcal{N}=4 $ SYM differs significantly from that of the non commutative $ \mathcal{N}=4 $ SYM as long as we consider the hydrodynamic description of both the theories with respect to their shear channels. However, the comparison remains incomplete as the analysis of the diffusive modes, in particular the computation of the $ R $ charge diffusion corresponding to non commutative $ \mathcal{N}=4 $ SYM  theory is still lacking in the literature. The purpose of the present article is therefore to fill up this gap and make a systematic comparison between two different anisotropic theories at strong coupling. In order to do that we essentially turn on $ U(1) $ fluctuations in the bulk and compute the corresponding $ R $ charge diffusion rates along both the commutative as well as the non commutative directions of the brane. Unlike the case for the shear viscosity \cite{Landsteiner:2007bd}, we observe a significant deviation in the charge transport phenomena along the non commutative direction of the brane. On the other hand, the charge diffusion constant along the direction of the commutative coordinates of the brane does not receive any non commutative corrections and thereby remains unchanged.

The organization of the paper is the following: In Section 2, we discuss the geometrical construction in the dual gravitational counterpart of the non commutative $ \mathcal{N}=4 $ SYM plasma. In Section 3, we explicitly compute the holographic charge diffusion rates both along the commutative as well as the non commutative directions of the brane and found that unlike the case for the shear modes their ratio is different from the unity. Finally, we conclude in Section 4.

%%%%%%%%%%%%%%%%%%%%%%%%%%%%%%%%%%%%%
\section{The dual set up}
We start our analysis with a formal introduction to the geometrical construction in the bulk space time that is holographically dual to non commutative $ \mathcal{N}=4 $ SYM theory at strong coupling. It is already known from the earlier literature that non commutative gauge theories at strong coupling could be consistently obtained from string theory by considering the so called decoupling limit in a system of $ Dp $ branes in the presence of a background NS B field that gives rise to certain scale of non commutativity in the large $ N $ limit \cite{Seiberg:1999vs}-\cite{Hashimoto:1999ut}. To start with, we consider the non commutative $ \mathcal{N}=4 $ SYM theory at finite temperature whose dual counterpart in the string frame reads as \cite{Landsteiner:2007bd},
\begin{eqnarray}
ds_{10}^{2}&=&\mathcal{H}^{-1/2}(-f dt^{2}+dx^{2}+h(dy^{2}+dz^{2}))+\mathcal{H}^{1/2}(f^{-1}dr^{2}+r^{2}d\Omega^{2}_{5})\nonumber\\
f&=& 1- \frac{r_H^{4}}{r^{4}},~~h=\frac{1}{1+\Theta^{2}\mathcal{H}^{-1}},~~\mathcal{H}=\frac{L^{4}}{r^{4}}
\end{eqnarray}
where, $ \Theta $ is the so called non commutative parameter and $ r_H $ is the usual position of the horizon. Following the AdS/CFT prescription, one could write $ L^{4}=4 \pi g^{2}_{YM}N \alpha'^{2} $ which in the decoupling ($ \alpha' \rightarrow 0 $) limit corresponds to a large value of $ N $ where $ N $ is the number of $ D3 $ branes. Finally, setting $ u = r_H^{2}/r^{2} $ the effective five dimensional metric in the Einstein frame could be formally expressed as \cite{Landsteiner:2007bd}, 
\begin{eqnarray}
ds^{2}&=& h^{-1/4}\mathcal{H}^{-1/2}(-f dt^{2}+dx^{2}+h(dy^{2}+dz^{2}))+\frac{L^{2}h^{-1/4}}{4u^{2}f}du^{2}\nonumber\\
f(u)&=& 1-u^{2},~~h(u)=\frac{u^{2}}{u^{2}+a^{2}},~~\mathcal{H}(u)= \frac{u^{2}}{u_T^{2}},~~u_T = \frac{r_H^{2}}{L^{2}},~~a=\Theta ~ u_T .\label{E1}
\end{eqnarray}
 
 Eq (\ref{E1}) is in fact the starting point of our analysis. In the above mentioned coordinate system (\ref{E1}) the horizon is placed at $ u=1 $ and the boundary is located at $ u=0 $. One should take a note on the fact that here $ (t,x) $ are the usual commutative directions whereas on the other hand, the other two spatial coordinates $ (y,z) $ exhibit the non commutative nature \cite{Landsteiner:2007bd}. 
From (\ref{E1}), it is in fact quite evident that due to presence of the non commutativity along two of the spatial directions of the brane, the full $ SO(3) $ symmetry of the boundary theory is reduced down to $ SO(2) $ leaving the rotational invariance only over the $ (y-z) $ plane of the brane. Finally, from (\ref{E1}) it is in fact quite trivial to note down the corresponding Hawking temperature which for the present case turns out to be,
\begin{eqnarray}
T = \frac{1}{\pi u_T L}.
\end{eqnarray}

%%%%%%%%%%%%%%%%%%%%%%%%%%%%%%%%%%%%%%%%%%%%%%%%%%%%%%%%%%
\section{Charge diffusion}
Based on the original prescription \cite{Policastro:2002se}-\cite{Kovtun:2008kx} for evaluating retarded Green's function corresponding to $ U(1) $ currents ($ J_{\mu} $), the purpose of the present section is to first make a systematic analytic investigation of the DC conductivity ($ \sigma_{DC} $) both along the commutative as well as the non commutative directions of the brane and then compute the corresponding R- charge diffusion(s) ($ \mathfrak{D} $) using the so called Einstein relation, $ \mathfrak{D}=\sigma_{DC}/ \chi $, where $ \chi $ is the charge susceptibility and $ \sigma_{DC} $ is the DC electrical conductivity that could be formally expressed as \cite{Policastro:2002se}-\cite{Kovtun:2008kx},
\begin{eqnarray}
\sigma_{DC} &=& - \lim_{\mathfrak{w} \rightarrow 0}\frac{1}{\mathfrak{w}}Im\ \mathcal{G}^{R}_{ii}(\mathfrak{w} , \mathfrak{q}=0)\nonumber\\
\mathcal{G}^{R}_{ii}(\mathfrak{w} , \mathfrak{q}=0) &=& -i \int d\tau\ d\textbf{x}\ e^{i\mathfrak{w} \tau}\ \Delta (t)\ \langle[J_i (\textbf{x}), J_i (0)]\rangle. \label{E13}
\end{eqnarray}

In order to compute the above quantity in (\ref{E13}) and thereby the charge diffusion ($ \mathfrak{D} $), we essentially study the dynamics of vector $ U(1) $ perturbations over the fixed back ground of the anisotropic black brane (\ref{E1}) \cite{Policastro:2002se}-\cite{Kovtun:2008kx}. Dynamics of these vector perturbations are in general governed by the Maxwell's action namely,
\begin{eqnarray}
S_M = -\frac{1}{4g^{2}_{M}}\int d^{5}x \sqrt{-g}\mathcal{F}_{ab}\mathcal{F}^{ab}\label{E14}
\end{eqnarray}
where, $ g^{2}_{M} $ stands for the Maxwell coupling of the $ U(1) $ theory.

The basic physics behind our analysis rests on the fact that the infra red behavior of these $ U(1) $ fluctuations in the bulk is solely governed by the \textit{hydrodynamics} where the dispersion relation of the type $ \mathfrak{w}=-i \mathfrak{D}\mathfrak{q}^{2} $ appears naturally as a consequence of the pole appearing in the Laplace transformed version of the charge density in the complex $ \mathfrak{w} $ plane which could be interpreted as a natural consequence of the diffusion of conserved charges.  In our analysis, considering the so called hydrodynamic limit namely, $ q \ll T $ we study fluctuations of the type, $\mathcal{A}_{m}\sim e^{i q.x}\mathcal{A}_{m}(t,u) $ over the background of (\ref{E1}). These fluctuations by means of the equation of motion as well as the relevant boundary conditions finally yield the dispersion relation of the above form in the limit $ \mathfrak{q}\rightarrow 0 $.

%%%%%%%%%%%%%%%%%%%%%%%%%%
\subsection{Charge susceptibility}
The purpose of the present section is to compute the charge susceptibility ($ \chi $) corresponding to non commutative $ \mathcal{N}=4 $ SYM plasma at strong coupling. In the AdS/CFT framework, the dual geometry corresponding to this non commutative plasma (at finite temperature) is essentially described by the five dimensional black hole solution (\ref{E1}) in the bulk space time.

In our computations, we strictly follow the methods proposed in \cite{Kovtun:2008kx}. The bottom line of our analysis is the following: In order to compute the susceptibility ($ \chi $), one needs to systematically solve the temporal gauge field ($ \mathcal{A}_{t} $) in the bulk consistent with the boundary condition at the horizon ($ u=1 $). 

The Maxwell equation that directly follows from (\ref{E14}) could be formally expressed as,
\begin{eqnarray}
\frac{1}{\sqrt{-g}}\partial_{\mu}(\sqrt{-g}\mathcal{F}^{\mu\nu})=0.\label{E15}
\end{eqnarray}

 The equation of motion corresponding to $ \mathcal{A}_{t} $ that readily follows from (\ref{E15}) could be formally expressed as,
\begin{eqnarray}
\mathcal{A}''_{t}+\frac{\partial_u(\sqrt{-g}g^{tt}g^{uu})}{\sqrt{-g}g^{tt}g^{uu}}\mathcal{A}'_{t}=0.\label{E16}
\end{eqnarray}
The corresponding solution turns out to be,
\begin{eqnarray}
\mathcal{A}_{t}(u)= \mathfrak{C}_{2}+\frac{4\mathfrak{C}_{1} \left(a^2+u^2\right)^{7/8} \left(7 u^2 \, _2F_1\left(1,\frac{3}{2};\frac{13}{8};-\frac{u^2}{a^2}\right)-5 a^2\right)}{15 a^2 u^{3/4}}.\label{E7}
\end{eqnarray}

The coefficient $\mathfrak{C}_{2}$ is uniquely determined by demanding the fact that $ \mathcal{A}_{t} $ must vanish at the horizon ($ u=1 $) which yields,
\begin{eqnarray}
\mathfrak{C}_{2}= \frac{4 \mathfrak{C}_{1} \left(a^2+1\right)^{7/8} \left(5 a^2-7 \, _2F_1\left(1,\frac{3}{2};\frac{13}{8};-\frac{1}{a^2}\right)\right)}{15 a^2}.
\end{eqnarray}
On the other hand, the chemical potential is given by,
\begin{eqnarray}
\mu = \mathcal{A}_{t}(u)|_{u\rightarrow \varepsilon}=\mathfrak{C}_{2} -\frac{4\mathfrak{C}_{1} a^{7/4} }{3 \varepsilon^{3/4}}+\mathcal{O}(\varepsilon^{5/4})\label{E9}
\end{eqnarray}
where $ |\varepsilon|\ll 1 $. Clearly the above quantity in (\ref{E9}) diverges in the limit $ \varepsilon \rightarrow 0 $. In order to have a finite chemical potential for the bondary theory, we thereby define the renormalized chemical potential as,
\begin{eqnarray}
\mu_{R}= \lim_{\varepsilon \rightarrow 0}\left(\mu +\frac{4 \varepsilon}{3} \frac{\partial \mu}{\partial \varepsilon} \right)= \mathfrak{C}_{2}.\label{E10}
\end{eqnarray}

Finally, using (\ref{E7}) the charge density could be readily obtained as \cite{Kovtun:2008kx},
\begin{eqnarray}
\varrho = \frac{\delta S_{M}}{\delta \mathcal{A}_{t}}|_{u =0}= \frac{2 u_T \mathfrak{C}_{1}}{g^{2}_{M}L}.\label{E22}
\end{eqnarray}
  
Using (\ref{E10}) and (\ref{E22}), the charge susceptibility finally turns out to be,
\begin{eqnarray}
\chi = \frac{\varrho}{\mu_{R}}=\frac{15 a^{2}u_T}{2 g^{2}_{M}L \left(a^2+1\right)^{7/8} \left(5 a^2-7 \, _2F_1\left(1,\frac{3}{2};\frac{13}{8};-\frac{1}{a^2}\right)\right)}\approx \frac{u_{T}}{2g^{2}_{M}L}.\label{E12}
\end{eqnarray}
Interestingly here we note that the charge susceptibility ($ \chi $) (almost) does not get corrected in the non commutative parameter ($ a $) upto fifth orders in the perturbation series. 

Having done these computations on charge susceptibility ($ \chi $), our next task would be to compute DC conductivities along both the commutative as well as non commutative directions of the brane. We denote $ \sigma_{\perp} $ as the conductivity along the commutative direction of the brane and $ \sigma_{\parallel} $ as the conductivity along the non commutative direction of the brane. Our purpose is to make a systematic comparison between these two conductivities and compare our results with the already existing results in the context of anisotropy \cite{Rebhan:2011vd}.

%%%%%%%%%%%%%%%%%%%%%%%%%%%%%%%%%%%%%%%%%%%%%%%%%%%%%%%%%%%%
\subsection{Conductivity I: $ \sigma_{\perp} $}
As a first part of our analysis, we compute the DC electrical conductivity along one of the commutative directions of the brane, namely the $ x $ direction.
We consider fluctuations of the form,
\begin{eqnarray}
\mathcal{A}_{m}(u,t)=L \int d\mathfrak{w}e^{-i \mathfrak{w}t}\mathcal{A}_{m}(u).\label{E24}
\end{eqnarray}

Considering $ m =x $ and substituting (\ref{E24}) into (\ref{E15}), we obtain
\begin{eqnarray}
\mathcal{A}_{x}'' +\frac{\partial_u (\sqrt{-g}g^{uu}g^{xx})}{\sqrt{-g}g^{uu}g^{xx}}\mathcal{A}_{x}' -\mathfrak{w}^{2}\frac{g^{tt}}{g^{uu}}\mathcal{A}_{x}=0.\label{E25}
\end{eqnarray}

In order to solve the above equation (\ref{E25}) in the so called low frequency regime, we chose the following ansatz, namely,
\begin{eqnarray}
\mathcal{A}_{x}=(1-u)^{\alpha} \Psi (u).\label{E26}
\end{eqnarray}

Considering the so called in going wave boundary condition \cite{Policastro:2002se}-\cite{Kovtun:2008kx}, our first task is to explore the above equation (\ref{E25}) in the near horizon limit of the brane namely, $ u \sim 1 $. This essentially enable us to determine the coefficient $ \alpha $ uniquely. Substituting (\ref{E26}) into (\ref{E25}) and considering the incoming wave boundary condition near the horizon of the black brane it is in fact quite trivial to show that,
\begin{eqnarray}
\alpha = - \frac{i \mathfrak{w}}{4 \pi u_T^{3/2} T}.
\end{eqnarray}

Our next task would be to substitute (\ref{E26}) in to (\ref{E25}) and solve $ \Psi (u) $ perturbatively in the frequency $ \mathfrak{w} $ near the boundary of the space time. This will finally enable us to compute the DC conductivity ($ \sigma_{\perp} $). In order to solve $ \Psi (u) $ perturbatively in the frequency ($ \mathfrak{w} $) we consider the following perturbative expansion namely,
\begin{eqnarray}
\Psi (u)= \Psi ^{(0)} + i(\mathfrak{w}/T)\Psi^{(1)}+\mathcal{O}(\mathfrak{w}^{2}/T^{2})
\end{eqnarray}
where each of the individual coefficients satisfies equation of motion of the following form,
\begin{eqnarray}
\Psi''^{(1)}+\frac{1}{2 \pi u_T^{3/2} (1-u)}\left[ \Psi'^{(0)}+\frac{1}{2(1-u)}\Psi^{(0)}\right] +\frac{\partial_u (\sqrt{-g}g^{uu}g^{xx})}{\sqrt{-g}g^{uu}g^{xx}}\left[\Psi'^{(1)}+\frac{1}{4 \pi u_T^{3/2} (1-u)}\Psi^{(0)} \right] &=&0\nonumber\\
\Psi''^{(0)}+\frac{\partial_u (\sqrt{-g}g^{uu}g^{xx})}{\sqrt{-g}g^{uu}g^{xx}}\Psi'^{(0)}&=&0.\label{E18}\nonumber\\
\end{eqnarray}

In the following we quote corresponding solutions one by one. Let us first consider the second equation in (\ref{E18}). The corresponding solution turns out to be,
\begin{eqnarray}
\Psi^{(0)}= \frac{4 \mathfrak{d}_{1}\mathfrak{Z}}{195 u^{3/4} \sqrt[8]{a^2+u^2}}+\mathfrak{d}_{2} \label{e19}
\end{eqnarray}
where,
\begin{eqnarray}
\mathfrak{Z}=  u^2 \sqrt[8]{\frac{u^2}{a^2}+1} \left(13 \left(3 a^2+7\right) F_1\left(\frac{5}{8};\frac{1}{8},1;\frac{13}{8};-\frac{u^2}{a^2},u^2\right)
-20 u^2 F_1\left(\frac{13}{8};\frac{1}{8},1;\frac{21}{8};-\frac{u^2}{a^2},u^2\right)\right)
\nonumber\\
-65 \left(a^2+u^2\right). \label{E30}
\end{eqnarray}
 In the above, we have expressed solution (\ref{e19}) in terms of Appell polynomials where the coefficients $ \mathfrak{d}_{1} $ and $ \mathfrak{d}_{2} $ are related to each other through the condition $ \Psi^{(0)}(1)=0 $. On top of it, one can also impose the asymptotic normalization condition which for the present case turns out to be, $ \Psi^{(0)}(0)=1/L $. These two conditions should in principle sufficient to determine these unknown coefficients uniquely. However, for the present purpose of our analysis it is sufficient to know the boundary behaviour of the gauge fields since we will be finally evaluating the entities near the boundary of the space time. Expanding (\ref{E30}) near the boundary ($ u \sim 0 $) of the space time we note, 
\begin{eqnarray}
\Psi^{(0)} \approx \frac{1}{L}\left( 1 - \frac{4 a^{7/4}}{3 \varepsilon^{3/4}}\right)^{-1}  \left( 1 - \frac{4 a^{7/4}}{3 u^{3/4}}+\frac{\left(8 a^2+7\right) u^{5/4}}{10 \sqrt[8]{a^2}}\right)+\mathcal{O}(u^{13/4})\label{E20}
\end{eqnarray}
where the numerical prefactor guarantees a normalized mode at the boundary. Note that here $ \varepsilon $ is the UV cut off as mentioned earlier. At the end of our calculations we finally consider the $ \varepsilon \rightarrow 0 $ limit in order to extract the finite piece at the boundary.

In the subsequent analysis we drop all the terms starting with quadratic order in $ u $.  Since $ u $ ranges between zero and one therefore it is indeed quite logical to truncate solutions upto certain order in $ u $ and particularly consider those terms that contribute significantly near the boundary of the space time. Using (\ref{E20}), the solution corresponding to ($ \Psi^{(1)} $) finally turns out to be,
\begin{eqnarray}
\Psi^{(1)} \approx \frac{1}{L}\left( 1 - \frac{4}{3 \varepsilon^{3/4}}\right)^{-1} \left( 1 -\frac{4 }{3 u^{3/4}}+\frac{a^{7/4}  \sqrt[4]{u}}{3 \pi  u_T^{3/2}}+\frac{a^{7/4} u^{5/4}}{6 \pi  u_T^{3/2}}-\frac{u}{4 \pi  u_T^{3/2}}\right) + \mathcal{O}(u^{2}).\label{E21}
\end{eqnarray}

Using (\ref{E20}) and (\ref{E21}), the non trivial piece in the DC conductivity (along $ x $- direction) finally turns out to be,
\begin{eqnarray}
\sigma_{\perp}= \frac{u_T}{g^{2}_{M}L T}.\label{eq23}
\end{eqnarray}
Finally, from (\ref{E12}) and (\ref{eq23}) one can easily read of the corresponding charge diffusion coefficient as,
\begin{eqnarray}
\mathfrak{D}_{\bot}=\sigma_{\perp}/\chi \sim \frac{1}{T}\label{E23}
\end{eqnarray}
where we have ignored the over all numerical pre factor. The above result (\ref{E23}) also follows from simple dimensional arguments. For example, it is straightforward to notice from 
(\ref{E12}) that $ [\chi]=1/L^{2} $ since the dimension of the Maxwell coupling in five dimensions goes as $ [g^{2}_{M}]=L $ \cite{Kovtun:2008kx}. On the other hand, following the same line of arguments we note $ [\sigma_{\perp}]=1/L^{2} $. Using these facts it could be readily seen that $ [\mathfrak{D}_{\bot}]=L $, where we have used the fact $ [T]=1/L $.

Eq.(\ref{eq23}) is in fact an important observation in itself. It reveals certain important fact that the DC conductivity ($ \sigma_{\perp} $) along the commutative direction of the brane does not get modified due to the presence of the non commutative parameter ($ \Theta $). The same line of argument also holds for the corresponding charge diffusion rate ($ \mathfrak{D}_{\bot} $).

%%%%%%%%%%%%%%%%%%%%%%%%%%%%%
\subsection{Conductivity II: $ \sigma_{\parallel} $}
For the sake of completeness as well as the clarity, our final task would be to compute the DC conductivity along one of the non commutative directions of the brane, say the $ y $ direction and make a systematic comparison of our results  with the result obtained in the previous section. To do that we first turn on fluctuations of the type,
\begin{eqnarray}
\mathcal{A}_{y}(u,t)=L \int d\mathfrak{w}e^{-i \mathfrak{w}t}\mathcal{A}_{y}(u)
\end{eqnarray}
which satisfy differential equation of the following form,
\begin{eqnarray}
\mathcal{A}_{y}'' +\frac{\partial_u (\sqrt{-g}g^{uu}g^{yy})}{\sqrt{-g}g^{uu}g^{yy}}\mathcal{A}_{y}' -\mathfrak{w}^{2}\frac{g^{tt}}{g^{uu}}\mathcal{A}_{y}=0.\label{e25}
\end{eqnarray}

To solve (\ref{e25}), we choose the following ansatz,
\begin{eqnarray}
A_{y}=(1-u)^{\beta}\Phi(u)
\end{eqnarray}
where the coefficient $ \beta $ could be readily obtained from the near horizon data namely, 
\begin{eqnarray}
\beta = - \frac{i \mathfrak{w}}{4 \pi u_T^{3/2} T}.
\end{eqnarray}

Like in the previous case, the function $ \Phi(u) $ could be solved perturbatively in the frequency $ \mathfrak{w} $ which in the hydrodynamic limit ($\mathfrak{w}/T \ll 1 $) yields the following set of equations namely,
\begin{eqnarray}
\Phi''^{(1)}+\frac{1}{2 \pi u_T^{3/2} (1-u)}\left[ \Phi'^{(0)}+\frac{1}{2(1-u)}\Phi^{(0)}\right] +\frac{\partial_u (\sqrt{-g}g^{uu}g^{yy})}{\sqrt{-g}g^{uu}g^{yy}}\left[\Phi'^{(1)}+\frac{1}{4 \pi u_T^{3/2} (1-u)}\Phi^{(0)} \right] &=&0\nonumber\\
\Phi''^{(0)}+\frac{\partial_u (\sqrt{-g}g^{uu}g^{yy})}{\sqrt{-g}g^{uu}g^{yy}}\Phi'^{(0)}&=&0.\label{e28}\nonumber\\
\end{eqnarray}
The corresponding solutions turn out to be,
\begin{eqnarray}
\Phi^{(0)} &= & \frac{1}{L}\left[1+ \frac{4 u^{5/4} \sqrt[8]{\frac{u^2}{a^2}+1} F_1\left(\frac{5}{8};\frac{1}{8},1;\frac{13}{8};-\frac{u^2}{a^2},u^2\right)}{5 \sqrt[8]{a^2+u^2}}\right] \nonumber\\
& \approx &\frac{1}{L}\left[ 1+\frac{4  u^{5/4}}{5 \sqrt[8]{a^2}}\right]+\mathcal{O}(u^{13/4}) \nonumber\\
\Phi^{(1)} & \approx & \frac{1}{L}\left[ 1+\frac{4u^{5/4} }{5}-\frac{ u}{4 \pi  u_T^{3/2}} \right]+\mathcal{O}(u^{2}).\label{e29}
\end{eqnarray}

Using (\ref{e29}), the corresponding DC conductivity finally turns out to be,
\begin{eqnarray}
\sigma_{\parallel}=\frac{u_T}{g^{2}_{M}LT}(1-\Theta^{1/4}u_T^{1/4}).\label{e31}
\end{eqnarray}

The way one would like to interpret the above result (\ref{e31}) is essentially the following: Unlike the previous case, the the DC conductivity ($ \sigma_{\parallel} $) along the non commutative direction of the brane is modified due to the presence of the non commutative parameter, and most importantly, the non commutative effects essentially suppress the value of the conductivity from that of its usual value corresponding to the commutative case. The same arguments also hold for the corresponding charge diffusion ($ \mathfrak{D}_{\parallel} $).

 Finally, the ratio between the two charge diffusion rates turn out to be,
\begin{eqnarray}
\frac{\mathfrak{D}_{\parallel}}{\mathfrak{D}_{\perp}} =\frac{\sigma_{\parallel}}{\sigma_{\perp}}= 1-\Theta^{1/4}u_T^{1/4}.\label{e32}
\end{eqnarray}
Eq.(\ref{e32}) is the full non perturbative result in the non commutative parameter ($ \Theta $) and is consistent with the corresponding result in the commutative ($ \Theta\rightarrow 0 $) limit. The crucial observation that one should make at this stage is the fact that unlike the case for the shear viscosity to entropy ($ \eta/s $) ratio \cite{Landsteiner:2007bd}, the charge diffusion rates are rather different along different directions of the brane. In other words, the charge diffusion is sensitive to the intrinsic anisotropy of the plasma. Finally, before we conclude, it is important to emphasis that similar observations have also been made earlier in a different context of anisotropy where people had observed, $ \sigma_{anisotropy}\neq \sigma_{isotropy} $ \cite{Rebhan:2011vd}.

%%%%%%%%%%%%%%%%%%%%%%%%%%%%%%%%%%
\section{Summary and final remarks}
Let us now summarize the key findings of our analysis. In our analysis, considering the so called hydrodynamic limit, we explore the charge transport phenomena in non commutative $ \mathcal{N}=4 $ SYM plasma at strong coupling. The motivation of our current analysis rests on the earlier results on shear viscosity to entropy ($ \eta/s $) ratio which was found to be universal despite of the intrinsic anisotropy of the $ Dp $ brane \cite{Landsteiner:2007bd}. In our analysis, however we observe that unlike the case for the $ \eta/s $ ratio, the charge diffusion rates are indeed different along two different directions of the brane. In particular, we observe that the holographic DC conductivity gets significantly modified (\textit{only}) along the non commutative directions of the brane and its value is in fact turns out to be lower compared to its commutative counterpart. Therefore we might conclude that from the point of view of the charge transport property, both the $ \theta $ deformed as well as the non commutative $ \mathcal{N}=4 $ SYM theories exhibit some sort of similarity whereas on the other hand, they differ quite significantly when we compare them with respect to their shear channels. Finally, it is noteworthy to mention that our result smoothly matches to that with the corresponding commutative result in the limit of vanishing $ \Theta $. 
\\ \\\
%%%%%%%%%%%%%%%%%%%%%%%%%%%%%%%%%%%%%%%%%%%%%%%%%%%%%%%%%%%%%%%%%%%%%%
{\bf {Acknowledgements :}}
 The author would like to acknowledge the financial support from CHEP, Indian Institute of Science, Bangalore.\\

%%%%%%%%%%%%%%%%%%%%%%%%%%%%%%%%%%%%%%%%%%%%%%%%%%%%%%%%%%%%%%%%%%%%%%%%%%%%%%%%%%%

\end{document}